\documentclass{ifacconf}

\usepackage{graphicx}      
\usepackage{natbib}        
\usepackage{xcolor} 
\usepackage{amsmath}
\usepackage{algorithm}
\usepackage[noend]{algpseudocode}
\usepackage{booktabs}
\usepackage{balance}
\usepackage{bm}

\begin{document}
\begin{frontmatter}

\title{Online Aging-Aware Energy Optimization for Vehicle-Home-Grid Integration\thanksref{footnoteinfo}}

\thanks[footnoteinfo]{This work was supported by the Swedish Energy Agency within the Vehicle Strategic Research and Innovation Program (Grant No.~P2022-00960) and the European Union's Horizon Europe program under the Marie Skłodowska-Curie (Grant No.~101131278).}

\author[First]{Francesco Popolizio} 
\author[First]{Torsten Wik} 
\author[Second]{Chih Feng Lee}
\author[First]{Changfu Zou}

\address[First]{Chalmers University of Technology, Göteborg, Sweden \\(e-mail: \{frapop, torsten.wik, changfu.zou\}@chalmers.se).}
\address[Second]{Polestar Performance AB, Göteborg, Sweden \\(e-mail: chih.feng.lee@chalmers.se).}

\begin{abstract}             
This paper investigates the economic impact of vehicle-home-grid integration through an online optimization algorithm that manages energy flows between an electric vehicle, a household, and the electrical grid.
The algorithm exploits vehicle-to-home (V2H) for self-consumption and vehicle-to-grid (V2G) for energy trading, adapting in real-time via a hybrid long short-term memory (LSTM) network for household load prediction and a nonlinear battery degradation model including cycle and calendar aging.
Simulations show annual economic benefits up to €3046.81 compared to smart unidirectional charging, despite a modest 1.96\% increase in battery aging.
Even under unfavorable market conditions, with no V2G revenue, V2H alone provides yearly savings of €425.48.
Sensitivity analyses on battery capacity, household load, and price ratios confirm the consistent benefits of bidirectional energy exchange, highlighting the role of EVs as active energy nodes for sustainable management.
\end{abstract}


\end{frontmatter}

\section{INTRODUCTION}
\label{sec:Introduction}
In recent years, the widespread adoption of electric vehicles (EVs) has experienced a significant growth, 
bringing major changes to the transportation sector.
EVs are providing a more sustainable solution to the challenges posed by climate change. However, this transition is also having a significant impact on global energy demand. With the exponential growth in the number of EVs on the road, the need for energy to recharge their batteries increases daily, placing significant challenges on electric grids.

The primary goal of traditional charging systems is to charge the battery once an EV is connected to the power socket. However, a vehicle is parked approximately 96\% of the time (\cite{2016EVparked96percent}). This highlights the potential to leverage EVs for active interaction with electric grids.  With the emergence of bidirectional power transfer, it is possible to reduce the costs for EV owners while supporting grid services. In this context, concepts such as vehicle-to-grid (V2G) and vehicle-to-home (V2H) are redefining EVs from mere transportation means into active energy nodes. V2G allows EVs to supply energy back to the grid, providing services such as load balancing and frequency regulation, while allowing the EV owner to generate profit. V2H allows EVs to supply power to a home, supporting home energy management, reducing costs and enhancing self-sufficiency (\cite{islam2022}).


Motivated by these opportunities, recent research has focused on optimizing energy exchanges between EVs, households, and the grid.
In the relevant literature, most approaches rely on offline optimization. For example, a mixed-integer linear programming (MILP) framework was developed by \cite{2015v2hv2g} to optimize the operation of smart households using V2G and V2H, aiming to minimize user costs. 
V2G and V2H applications for energy trading were also explored by \cite{2022v2hv2gGermany}, which reports that an average German household with a photovoltaic system, heat pump, and stationary battery can generate annual revenues of approximately €310. 
However, these two works ignored battery degradation in the optimization for simplicity, which tend to result in suboptimal solutions due to excessive battery costs. To mitigate the risk, \cite{2024milpChalmers} presented a MILP-based optimal scheduling model for EVs, in which V2G and a linearized battery degradation model were considered. Following this trend, \cite{2023Lee} presented a  linear programming-based V2G optimization focusing on frequency regulation according to the rules of the Swedish market.
A common feature of these works is that they assume full knowledge of input data in advance, and then solve the problem offline, making them less applicable to real-world with dynamic scenarios.

To mitigate the potential drawbacks of offline optimization, recent research has increasingly focused on optimizing V2G technology online.
\cite{2022onlineV2GPengfei} proposed an online V2G scheduling method based on fuzzy logic control to mitigate battery aging through a two-stage optimization. Cycle aging was considered only during offline calibration of the fuzzy parameters, while calendar aging was not explicitly accounted for during the online operation.
\cite{2024OnlineV2GPSO} used a particle swarm optimization to solve the V2G optimization with a sliding window; as for the battery degradation, it only considers the cycle aging with the rain-flow cycle counting method. 
Although battery degradation has been considered, the existing V2G optimization algorithms heavily rely overly simplified degradation models that fail to capture calendar and cycle aging comprehensively. Furthermore, these online algorithms have not systematically synergized V2G, V2H, battery dynamics, and the inherent uncertainties associated with household energy consumption. 

To address the identified research gaps, this paper proposes a nonlinear online optimization algorithm for vehicle-home-grid integration, aiming to minimize user costs through energy trading based on hourly price variations. The framework considers a single user owning both an EV and a household.
To operate online, the algorithm employs a hybrid long short-term memory (LSTM) neural network to forecast the household load, while the energy scheduling is determined through an online optimization procedure inspired by shrinking-horizon model predictive control (MPC).
A detailed battery model is incorporated to capture both calendar and cycle aging, accounting for their nonlinear dependence on coupled stress factors. As a result, the proposed method minimizes overall electricity expenses and battery degradation costs, contributing to sustainable and economically efficient energy management.

\section{METHODOLOGY}
\label{sec:2ch}
\begin{figure}
    \centering
    \includegraphics[width=0.9\linewidth]{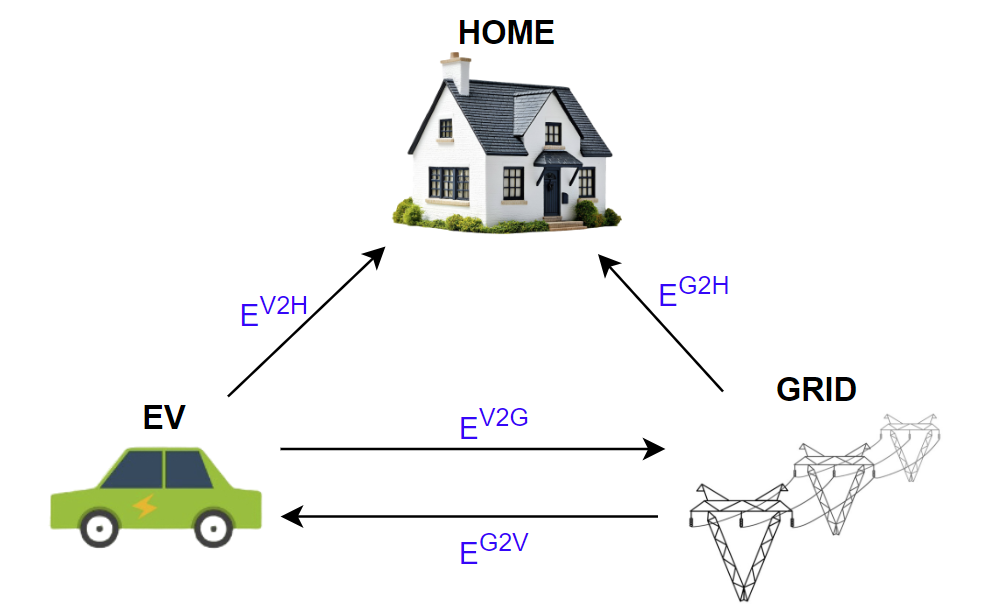}
    \caption{Proposed vehicle-home-grid integration framework. }
    \label{fig:actors}
\end{figure}

The considered vehicle-home-grid integration, as illustrated in Fig.~\ref{fig:actors}, has three actors: an EV, a house, and the power grid. The EV can supply energy to the grid and the house (V2G and V2H), while grid-to-vehicle (G2V) represents the energy used to recharge the EV. The house, on the other hand, can also receive energy in a standard way, directly from the power grid through grid-to-home (G2H). Therefore, the present work assumes that the EV is only charged when parked at home.

Since the model is implemented online, the exact time when the user parks the EV is not known in advance. However, once the EV is parked, the user communicates the time at which they will pick it up again.

The electricity price is known in advance, thanks to the day-ahead pricing. 
However, the household load is only known up to the current time, while future values remain uncertain.
To address this uncertainty, a hybrid LSTM neural network is used to predict the future household load. Additionally, at each time step the controller updates the predicted values using the newly available data.

The control strategy is formulated as a shrinking-horizon MPC problem, activated whenever the EV is parked. The prediction horizon corresponds to the parking duration, known only at the beginning of each parking session.

\subsection{Vehicle-Home-Grid Control}
While the EV is parked, an optimization problem is formulated to minimize the user cost for energy trading and battery degradation:
\begin{equation} \label{eq:obj}
    \min \sum_t EC_t + BC_t, 
\end{equation}
where $t$ represents time, and $EC_t$ and $BC_t$ denote the energy and battery costs, respectively. 

$EC_t$ and $BC_t$ can be calculated by
\begin{align}    
    EC_t =\:& (E^{G2V}_t + E^{G2H}_t) \cdot p_t - E^{V2G}_t \cdot \gamma \cdot p_t,  
    \label{eq:EC}
    \\ 
    BC_t =\:& NV \cdot \frac{BD_t(\%)}{100\% - EoL(\%)}, 
\label{eq:BC}
\end{align}
where the energy flows are expressed in kWh, and $p_t$ is the day-ahead energy price, expressed in €/kWh. 
Energy costs arise from purchasing energy from the grid for G2V and G2H, minus the profits from selling energy to the grid (V2G).
$\gamma$ is the price ratio, representing the ratio between the selling price and the buying price of energy. If $\gamma=1$, it means the buying and selling prices are equal. 
In \eqref{eq:BC}, $BD_t$ is the battery degradation in percentage of the initial battery capacity,
$EoL$ is the battery capacity at the end of life, which is assumed to be 80\%, and $NV$ is the net value of the battery.

Based on economic principles discussed in \cite{park2019_economics}, $NV$ is calculated through
\begin{equation}    \label{eq:NV}
    NV = C_{rep} \cdot \frac{1}{(1+i_r)^L} - C_{rv} \cdot \frac{1}{(1+i_r)^L},
\end{equation}
where $C_{rep}$ is the battery replacement cost, $C_{rv}$ is the battery residual value, $i_r$ is the yearly discount rate, and $L$ represents the nominal battery life in years.

The objective function in \eqref{eq:obj} is subject to the following set of constraints:
\begin{align}    
    E^{V2G}_t, E^{V2H}_t, E^{G2V}_t, E^{G2H}_t \geq \:& 0, \label{eq:c1}
    \\
        E^{G2V}_t \leq \:&  E_{\max}, \label{eq:c2}
    \\
        E^{V2G}_t + E^{V2H}_t \leq \:& E_{\max}, \label{eq:c3}
    \\
    0\% \leq SoC_t \leq \:& 100\%, \label{eq:c4}
    \\
    SoC_t = SoC_{t-1} + \frac{E^{G2V}_t}{E_b} - \: \frac{E^{V2G}_t}{E_b} 
    & - \frac{E^{V2H}_t}{E_b} - \frac{d_t}{D_r}, \:\: \label{eq:c5}
    \\
        SoC_{t} \geq\:& SoC^{goal}_t,  \label{eq:c6}
    \\
    HL_t =\:& E^{G2H}_t + E^{V2H}_t, \label{eq:c7}
    \\
        E^{G2V}_t + E^{G2H}_t \leq\:& G_t.    \label{eq:c8}
\end{align}
In \eqref{eq:c1}, all energy flows must be non-negative. The input energy $E^{G2V}_t$ and the output energy from the EV ($E^{V2G}_t+E^{V2H}_t$) cannot exceed the maximum limit $E_{\max}$, imposed by the EV charger, resulting in the constraints formulated in \eqref{eq:c2} and \eqref{eq:c3}.
The constraint \eqref{eq:c4} ensures that the EV's SoC ranges from 0 to 100\%.
The SoC evolution in \eqref{eq:c5} depends on the previous SoC, the energy exchanged with the battery (normalized by its capacity, $E_b$), and the distance traveled (normalized by the battery range, $D_r$). Notice that, while the EV is parked, $d_t=0$.
\eqref{eq:c6} requires the SoC to be at least $SoC^{goal}_t$, which is 80\% at pickup times and zero otherwise.
The household load $HL_t$ is met by energy supplied either from the grid or the EV, according to \eqref{eq:c7}.
However, the total energy purchased from the grid cannot exceed the available supply $G_t$ (which is here assumed to be sufficiently large to always ensure the constraint \eqref{eq:c8} is never active).

Note that \eqref{eq:obj}--\eqref{eq:c8} apply only when the EV is parked. However, the model simulates a realistic scenario where the EV alternates between parking and driving throughout the simulation. During driving, no optimization is needed, as the EV is absent and all its energy flows are zero. In this phase, the SoC depends on the previous step and the distance traveled $d_t$, while the household load is fully supplied by the grid ($E_t^{G2H}$).

\subsection{Battery Model}

An empirical battery model, experimentally validated on real-world lithium-iron-phosphate cells, is employed to describe both calendar and cycle aging. The complete model details and parameters are provided by \cite{2018Schimpe}, while we present only the key equations relevant to the vehicle-home-grid control.

The calendar aging $BD_t^{cal}$ is a function of temperature $T$, SoC and time in hours, as computed by
\begin{equation}    \label{eq:BD_cal}
    BD_t^{cal} = \frac{K^{cal}_t(T,SoC)}{2\sqrt{t}}\Delta t,
\end{equation}
with $t$ being the cumulative time and $\Delta t$ the sample time, assumed to be one hour in this work. $K^{cal}_t(\cdot,\cdot)$ is a stress factor which depends on $T$ and SoC according to
\begin{align}    \label{eq:Kcal}
    K^{cal}_t(T&,SoC) = k_{cal,ref}\! \cdot\! \exp \left[ \frac{-E_a^{cal}}{R_g}\! \left( \!\frac{1}{T}\!-\!\frac{1}{T_{ref}} \right) \!\right] \nonumber \\
      &\cdot \!\left[ \!\exp\! \left( \frac{\alpha  F}{R_g}  \frac{U_{a,ref}\!-\!U_a(SoC_t)}{T_{ref}}  \!\right) \!+\!k_0 \!\right],
\end{align}
where $U_a$ is the anode open-circuit potential (computed as a function of SoC). 
We assume a constant $T$ over time, equal to 15°C.

The cycle aging depends on three different conditions: high temperature, low temperature, and low temperature with high SoC.
The cycle aging due to high temperature $BD_t^{cyc,hT}$ is computed as follows:
\begin{equation}    \label{eq:BD_cyc_hT}
    BD_t^{cyc,hT} = \frac{K^{cyc,hT}_t(T)}{2\sqrt{Q_{tot}}}\Delta Q_{tot},
\end{equation}
where $Q_{tot}$ is the cumulative total energy throughput (Ah) for charging and discharging, while $\Delta Q_{tot}$ is the instantaneous total energy throughput (in $\Delta t$). 
$Q_{tot}$ is obtained by summing the energy in input/output (Wh) over time and dividing by the voltage.
The stress factor $K^{cyc,hT}_t(\cdot)$ is a function of $T$ in the form
\begin{equation}    \label{eq:Kcyc_hT}
    K^{cyc,hT}_t\! = k_{cyc,hT,ref} \!\cdot \! \exp \! \left[ \! \frac{-E_{a,hT}^{cyc}}{R_g} \! \left( \! \frac{1}{T} \!- \! \frac{1}{T_{ref}} \! \right) \! \right].
\end{equation}
The cycle aging due to low temperature $BD_t^{cyc,lT}$ is computed by
\begin{equation}    \label{eq:BD_cyc_lT}
    BD_t^{cyc,lT} = \frac{K^{cyc,lT}_t(T,I_{Ch})}{2\sqrt{Q_{ch}}}\Delta Q_{ch}.
\end{equation}
Now, $Q_{ch}$ is the cumulative energy throughput (Ah) for charging. The stress factor $K^{cyc,lT}_t(\cdot,\cdot)$ is a function of $T$ and the charging current rate $I_{Ch}$ (obtained as $\Delta Q_{Ch}/\Delta t$), which is formulated as
\begin{align}    \label{eq:Kcyc_lT}
    K^{cyc,lT}_t  =  \:& k_{cyc,lT,ref}   \exp  \left[  \frac{-E_{a,lT}^{cyc}}{R_g}  \left(  \frac{1}{T} -  \frac{1}{T_{ref}}  \right) \right] \nonumber \\
    & \cdot \exp \left( \beta_{lT}  \frac{I_{Ch}  -  I_{Ch,ref}}{C_0}  \right).
\end{align}
The cycle aging due to low temperature and high SoC $BD_t^{cyc,lThSoC}$ is computed as
\begin{equation}    \label{eq:BD_cyc_lThSoC}
    BD_t^{cyc,lThSoC} = K^{cyc,lThSoC}_t(T, I_{Ch}, SoC) \cdot \Delta Q_{ch},
\end{equation}
where the stress factor $K^{cyc,lThSoC}_t(\cdot,\cdot,\cdot)$ is a function of $T$, current rate, and SoC, given by
\newcommand{\sign}{\operatorname{sgn}}      
\begin{equation}    \label{eq:Kcyc_lThSoC}
\begin{split}
     & K^{cyc,lThSoC}_t \!\! = \! k_{cyc,lThSoC,ref}   \exp \! \bigg[ \!  \frac{\!-E_{a,lThSoC}^{cyc}}{R_g}  \! \left( \! \frac{1}{T} \! - \! \frac{1}{T_{ref}} \! \right) \! \bigg] \! \\
     & \! \cdot \! \exp \! \left( \! \beta_{lThSoC} \frac{\!I_{Ch} \! - \! I_{Ch,ref}}{C_0} \! \right)  \frac{\sign\left(SoC_t \! - \! SoC_{ref}\right) \! + \! 1}{2}. 
\end{split}
\end{equation}

Finally, the total battery degradation $BD_t$ in \eqref{eq:BC}, defined as the capacity loss in percentage, is computed at each time step by
\begin{equation}    \label{eq:BD_tot}
    BD_t \! = \! BD_t^{cal} \! + \! BD_t^{cyc,hT} \! + \! BD_t^{cyc,lT} \! + \! BD_t^{cyc,lThSoC}.
\end{equation}

In this work, we assume that battery cells in the EV are well-managed, resulting in homogeneous characteristics across different cells. In accordance with a Polestar EV, we set the battery system's capacity $E_b$ to 82 kWh (corresponding to $C_0$ = 205 Ah), the nominal voltage to 400 V, and the vehicle's driving range $D_r$ as 514 km. Also, the maximum energy for charging/discharging ($E_{max}$) is set to 11 kWh.

\subsection{Household Load Prediction and Management}
In real-world applications, the household load is largely affected by weather and user behavior. Due to substantial uncertainties and stochasticity, the load values are typically unknown in advance. This poses challenges for the optimal energy flow to be computed when the EV is parked. 
To address this, we develop a real-time household load predictor based on historical data and patterns.
Specifically, a hybrid long short-term memory neural network (LSTM) is applied.
LSTM is very powerful in capturing dependencies in sequential data and is particularly suitable for time series energy consumption data that exhibit recurring daily and seasonal patterns.
The hybrid LSTM approach is adopted from (\cite{huang2021hybrid,2024_hybrid_lstm_changfu}) to incorporate additional features that influence energy consumption, providing the model with more contextual information and improved prediction accuracy.

The goal of the hybrid LSTM is to predict the household load one hour ahead ($t+1$). From a chosen dataset, a total of four features were extracted:
\begin{enumerate}
\renewcommand{\labelenumi}{$f\theenumi)$}
    \item energy consumption from the previous 23 hours up to the current time step (from $t-23$ to $t$), which serves as the input sequence for the LSTM neural network
\end{enumerate}
For each of these 24 hourly consumption values, three contextual features were included:
\begin{enumerate}
\setcounter{enumi}{1}
\renewcommand{\labelenumi}{$f\theenumi)$}
    \item the day of the year (ranging from 1 to 365)
    \item the day of the week (ranging from 0 to 6)
    \item the hour of the day (ranging from 0 to 23)
\end{enumerate}    
$f2, f3, f4$ (72 features in total) are fed into the dense neural network.

\begin{figure}[h]
    \centering
    \includegraphics[width=0.9\linewidth]{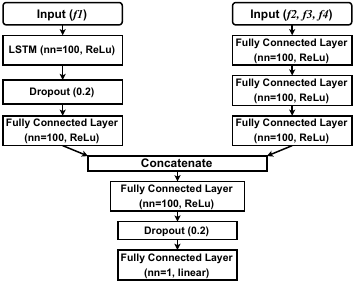}
    \caption{Hybrid LSTM neural network architecture, where nn labels the number of neurons.}
    \label{fig:NN_architecture}
\end{figure}

The architecture used, extensively studied in \cite{2024_hybrid_lstm_changfu}, is summarized in Fig.~\ref{fig:NN_architecture}.

Since our output is limited to a single hour of prediction, to predict multiple hours, we use the neural network recursively, where the output of each prediction becomes the input for the next. 

The hybrid LSTM neural network was implemented using Tensorflow and Keras. The model was trained with a batch size of 8 and for 75 epochs. 
The model employed the Adam optimizer.
The other settings are shown in Fig.~\ref{fig:NN_architecture}. 
As demonstrated in \cite{2024_hybrid_lstm_changfu}, this architecture shows accurate performances for electrical load prediction.

The proposed prediction framework operates during the time window in which the EV is parked.
During this interval, the household load is forecasted, and the predicted values are used as deterministic inputs to the optimization problem in \eqref{eq:obj}, which determines the energy scheduling for the upcoming hours. As time progresses, the actual household consumption becomes available, replacing the corresponding predicted values. The optimization is then re-solved with this updated information, effectively implementing a shrinking-horizon MPC scheme.

\subsection{Data Sources and Simulation Data Generation}
\begin{figure}[h]
    \centering
    \includegraphics[width=0.8\linewidth]{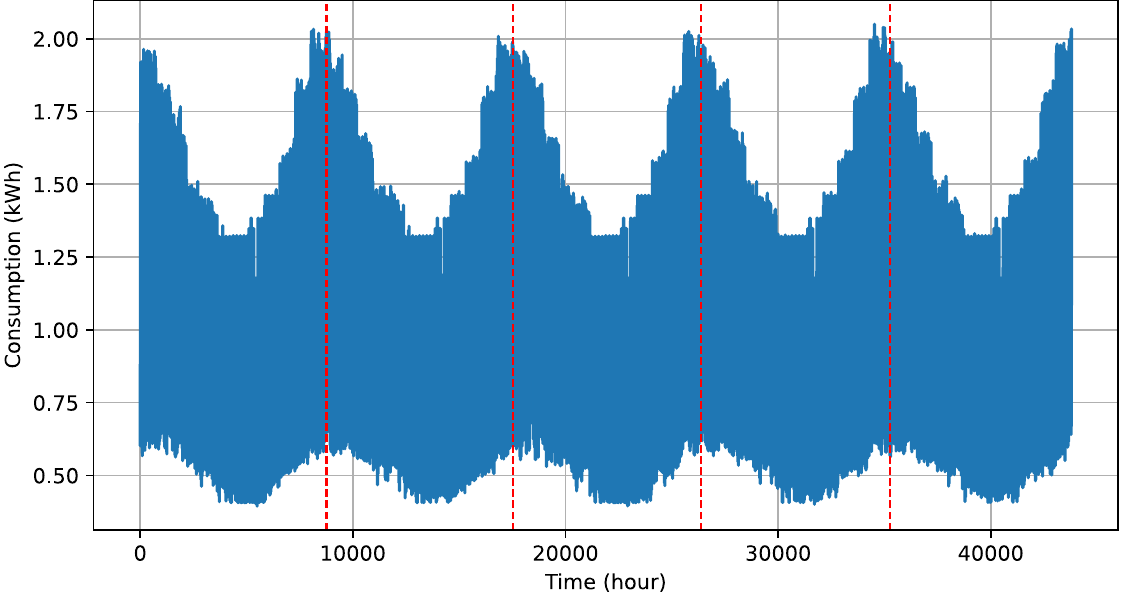}
    \caption{Concatenated household load for five apartments in the State of Washington, US. The red dashed vertical lines represent the transition between datasets.}
    \label{fig:HLdataset}
\end{figure}

For the household load, the dataset from \cite{2021_hl_data} has been used in this work. It contains household load data for single apartments in the US, each spanning one year. 
Five datasets from the state of Washington have been selected because they exhibit similar consumption patterns (shown in Fig.~\ref{fig:HLdataset}, where the datasets are concatenated to form a five-year dataset). 
Four of the five datasets were used for training, while one was used for testing, meaning 20\% of the data was reserved for the model simulation. 
This last year of data, dedicated for the simulations, has an average hourly use of 0.9 kWh, resulting in a daily average consumption of 21.6 kWh.

As for the battery, the replacement cost $C_{rep}$ is defined as 111.5 €/kWh, the residual value is $C_{rv} =30\% C_{rep}$, the yearly discount rate is $i_r= 10\%$ and the nominal battery life is $L=10$ years.

The daily EV usage and driving distances are modeled using a truncated Gaussian distribution. The EV is assumed to be picked up by the user between 6:00 and 10:00 am, with a mean pickup time of 8:00 am. 
The travel durations range from 7 to 11 hours, with a mean of 9 hours. 
The daily driving distance is assumed to lie between 30 and 40 km, with a mean of 35 km.
We further assume that the driving distance is distributed linearly over the duration of the travel session.

The electricity price for the year 2022 for Sweden (SE3 area) is taken from \cite{entsoe}. However, this represents the initial price ($p^{ini}$), i.e., the price without taxes. The final price $p$ used in the simulations has been calculated using
\begin{equation}
    p_t = p^{ini}_t \cdot (1+0.25) + 0.006.
\end{equation}
A 25\% markup is applied to the initial price to account for Swedish value-added tax (VAT), and a fixed cost of 0.006 €/kWh is added to represent the energy tax.

To match the resolution of the energy price and household load datasets, simulations are conducted over one year with an hourly time step. 
The simulation started with the battery with an age of 60 days and the EV's SoC at 60\%.
The complete control algorithm was developed in Python, with CasADi (\cite{casadi}) serving as the optimization framework and the IPOPT solver applied for solving the nonlinear optimization problems.

\section{RESULTS AND DISCUSSION}
\label{sec:3ch}
Simulations were conducted for two distinct scenarios to evaluate user costs in the vehicle-home-grid integration: 
\begin{enumerate}
    \renewcommand{\labelenumi}{\textit{\textbf{\Alph{enumi}.}}}
    \item Proposed vehicle-home-grid integration: With objective function in \eqref{eq:obj}, this scenario aims to minimize the user energy costs and battery degradation by controlling the bidirectional energy flows among the vehicle, household, and electric grid.
    \item Unidirectional smart charging (benchmark): This scenario seeks to minimize costs, including battery degradation, without employing V2G and V2H technologies. Therefore, the objective function in \eqref{eq:obj} is replaced by
    \begin{equation} \label{eq:obj_unidir}
        \min \sum_t (E^{G2V}_t + E^{G2H}_t) \cdot p_t + BC_t.
    \end{equation}
\end{enumerate}

\subsection{Cost and Battery Degradation Analysis}
With the price ratio $\gamma=1$, meaning the price for purchasing and selling energy is identical, Table~\ref{tab:1} summarizes the user costs for the two scenarios. Here, the user's final cost ($FC$) is calculated as the sum of the energy cost ($EC$) and battery degradation cost ($BC$).
$BD^{cyc}$ is the sum of cycle aging under high temperature, low temperature, and high temperature with low SoC. $E_{batt}$ represents the total energy that flows in and out of the EV battery. 

\setlength{\tabcolsep}{4.5pt} 
\begin{table}[h]
    \centering
    \caption{User costs and battery degradation for scenarios \textit{\textbf{A}} and \textit{\textbf{B}}.}
    \begin{tabular}{cccccccc}
        \toprule
                            & $FC$     & $EC$     & $BC$    & $BD$   & $BD^{cal} $  & $BD^{cyc}$ & $E_{batt}$      \\
                            & [€]      & [€]      & [€]     & [\%]   & [\%]         & [\%]       & [kWh]           \\
         \midrule
        \textbf{\textit{A}} & -1070.21 & -1739.38 & 669.17  & 5.42   & 2.26         & 3.16       & 45030      \\
        \textbf{\textit{B}} & 1976.60  & 1549.12  & 427.48  & 3.46   & 2.72         & 0.75       & 4054       \\
        \bottomrule
    \end{tabular}
    \label{tab:1}
\end{table} 

It can be observed that scenario \textbf{\textit{A}} yields favorable performance: 
the negative energy cost ($EC$) indicates profit generated by selling electricity back to the grid via V2G. 
Overall, scenario \textbf{\textit{A}} achieve an annual user profit of €1070.21. 

In contrast, scenario \textbf{\textit{B}} (benchmark) shows significantly lower battery energy flow ($E_{batt}$), limited to EV charging (G2V) and driving. Without energy sales (no V2G or V2H), scenario \textbf{\textit{B}} results in significantly higher energy costs and minimal battery degradation due to both optimized smart charging and reduced battery usage.

By comparing scenarios \textbf{\textit{A}} and \textbf{\textit{B}}, it can be concluded that vehicle-home-grid integration provides an annual economic advantage of €3046.81 for the user. Specifically, scenario \textbf{\textit{A}} degrades the battery by 1.96\% more but reduces the energy costs by €3288.50 compared to scenario \textit{B}.

\subsection{Sensitivity Analysis of Price Ratio $\gamma$}
\begin{figure}[ht]
    \centering
    \includegraphics[width=0.9\linewidth]{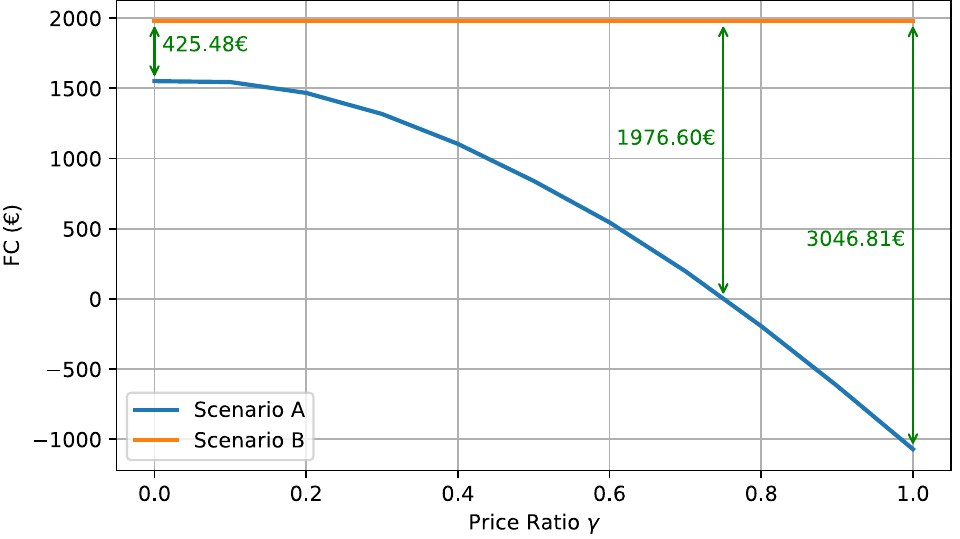}
    \caption{$FC$ for scenario \textbf{\textit{A}} and \textbf{\textit{B}} varying price ratio $\gamma$.} 
    \label{fig:cost_fee}
\end{figure}

The results above assume a price ratio $\gamma = 1$, commonly adopted as the most optimistic case. Namely, there is no price difference between the purchase and sale of energy.
Different values of $\gamma$ may significantly affect the final cost $FC$ due to changes in the energy cost $EC$. 
Fig.~\ref{fig:cost_fee} shows the final cost $FC$ for scenarios \textbf{\textit{A}} and \textbf{\textit{B}} across a range of $\gamma$ values from 0 to 1.

As expected, in scenario \textbf{\textit{B}}, the curve remains constant due to unidirectional smart charging, which does not involve V2G and thus is independent of the price ratio $\gamma$.
In scenario \textbf{\textit{A}}, for $\gamma=1$, $FC$ corresponds to the value shown in Table I. As $\gamma$ decreases, $FC$ increases because the reduced price difference between buying and selling makes V2G less profitable. 
For $\gamma=0.75$, $FC=0$. This indicates that using the EV for vehicle-home-grid integration eliminates EV-related costs, corresponding to an economic gain of €1976.60 that the user would have spent on unidirectional smart charging. 
However, for $\gamma<0.75$, $FC$ becomes positive, meaning that while the economic gain persists, no further profit is generated.
When $\gamma=0$, V2G is no longer performed as it offers no benefit; however, V2H continues, contributing to self-consumption. This results in an economic gain (corresponding to a saving) of €425.48 compared to scenario \textbf{\textit{B}}.

Overall, these results demonstrate that vehicle-home-grid integration consistently offers benefits: even in the worst-case scenario with $\gamma=0$, the user saves around €425 annually through bidirectional charging with their EV.

\subsection{Impact of Battery Size \& Household Load Variations}
\begin{figure}[h]
    \centering
    \includegraphics[width=0.9\linewidth]{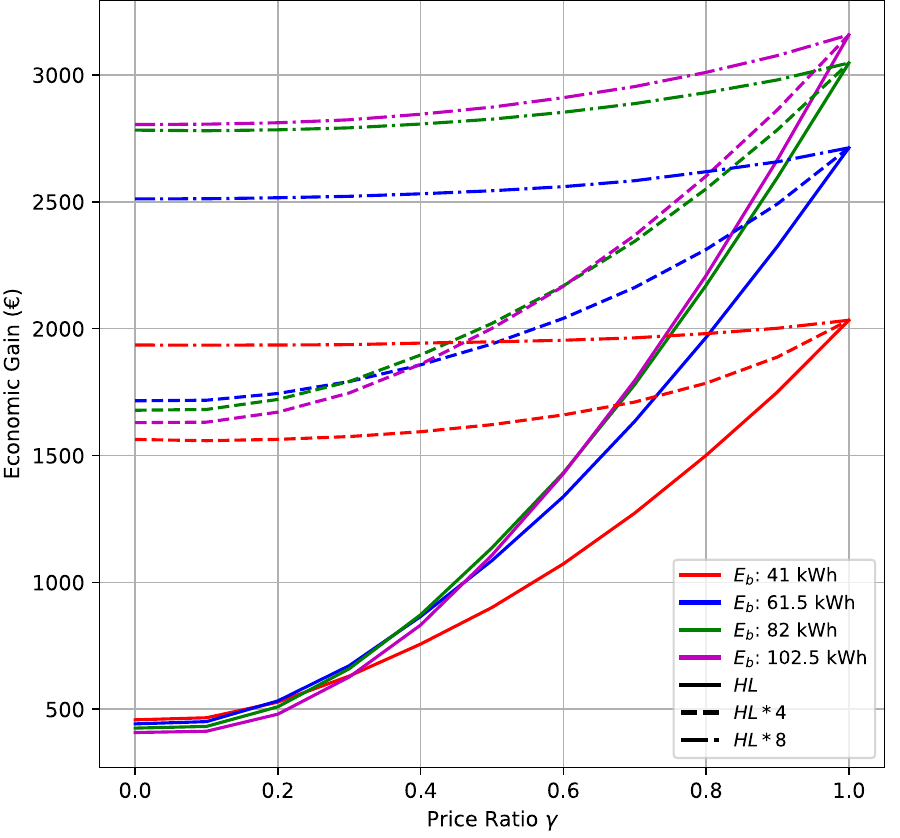}
    \caption{Economic gain for varying battery capacities ($E_b$), household loads, and price ratios ($\gamma$).}
    \label{fig:HL_bc_mix}
\end{figure}

Additional simulations are conducted to examine the influence of battery capacity ($E_b$ = [41, 61.5, \textbf{82}, 102.5] kWh) and household loads ($\bm{HL}$, $4HL$, $8HL$). 
Here, $HL$ is the consumption of a standard apartment, as shown in Fig.~\ref{fig:HLdataset}, while $4HL$ and $8HL$ correspond to larger households (86.4 and 172.8 kWh/day, respectively). 
Fig.~\ref{fig:HL_bc_mix} illustrates the economic gain, defined as the difference in $FC$ between scenario \textbf{\textit{B}} and \textbf{\textit{A}}, for varying $E_b$, household loads, and $\gamma$.

When the consumption is $HL$ (solid lines), the curves for different $E_b$ values exhibit a clear increasing slope with $\gamma$, due to more effective V2G utilization.  
At low $\gamma$ values (less than 0.2), the economic gain decreases with larger batteries because any of considered battery capacities is sufficient to cover this $HL$, whereas larger batteries lead to increased battery replace costs, i.e., $C_{rep}$ in \eqref{eq:NV}.  
Conversely, at higher $\gamma$ values, larger batteries result in greater economic gains due to enhanced V2G. 

With increased household consumption ($4HL$ and $8HL$), the economic gain curves become flatter as $\gamma$ increases, indicating reduced V2G opportunities due to the higher household demand. Specifically, for $4HL$ (dashed lines), a 41 kWh battery is clearly limited, giving the lowest economic gain among all tested battery sizes. Moreover, in the $4HL$ case, smaller batteries offer slightly higher gains when $\gamma$ is low, while the opposite occurs at higher $\gamma$, consistent with the behavior seen in the $HL$ case. For $8HL$ (dash-dot lines), the very high household load severely limits the availability of energy for grid export, suppressing V2G utilization. Unlike the $HL$ and $4HL$ cases, a larger battery consistently yields higher economic gain across the entire $\gamma$ range. This is because larger batteries provide the capacity needed to handle the high household demand more flexibly, even if V2G is minimally utilized.

Additionally, for $\gamma=1$ and the same $E_b$, the economic gain remains constant across household loads. 
This occurs because, with $\gamma=1$, there is no financial penalty on energy sales, and the user's benefit depends solely on battery capacity. As a result, while a higher household load increases $FC$ in scenarios \textbf{\textit{A}} and \textbf{\textit{B}}, their economic difference remains unchanged for the same battery size.

Overall, results in Fig.~\ref{fig:HL_bc_mix} confirm that vehicle-home-grid integration consistently provides economic benefits in all tested cases. 
Regardless of price ratios, battery capacities, and household loads, even in the least favorable conditions, the user achieves financial advantages compared to the unidirectional smart charging scenario.
\section{CONCLUSION}
\label{sec:4ch}
This paper proposed a real-time optimization algorithm for managing energy flows in vehicle–home–grid integration. The contributions first arise from the explicit consideration of nonlinear battery cycle and calendar aging, uncertainties in future household load, and constraints on vehicle usage and battery dynamics. By utilizing a hybrid LSTM neural network to predict household loads and a detailed battery model, this algorithm results in minimized user energy and battery costs, yielding an annual economic gain of up to €3046.81 for a single user over the benchmark scenario with unidirectional smart charging.   

The second major contribution is a systematic analysis of various price ratios, battery capacities, and household loads within the optimization problem. We have found that
\begin{itemize}
    \item Reduced price ratios make V2G less profitable. However, even without performing V2G, V2H contributes to self-consumption and gives an annual economic gain of €425.48. 
    \item For standard apartments and small to moderate-sized houses, larger batteries reduce the economic gains at low price ratios but provide greater benefits at high price ratios. 
    \item For large household loads, larger batteries always lead to higher economic gains, as more efficient household energy management is allowed.
\end{itemize}


\bibliography{ifacconf}

\end{document}